# Research on Model Parallelism and Data Parallelism Optimization Methods in Large Language Model–Based Recommendation Systems

Haowei Yang
Cullen College of Engineering
University of Houston
Houston, TX, USA
*Corresponding author:
hyang38@cougarnet.uh.edu

Yu Tian
Khoury College of Computer Science
Northeastern University
Seattle, WA, USA
tian.yu2@northeastern.edu

Zhongheng Yang
Khoury College of Computer Sciences
Northeastern University
Jersey City, NJ, USA
yang.zho@northeastern.edu

Zhao Wang
School of Computer Science and Big Data
Fuzhou University
Fuzhou, Fujian, China
2581461143@qq.com

Chengrui Zhou
Fu Foundation School of Engineering and Applied Science
Columbia University
New York, NY, USA
zhou.chengrui@columbia.edu

Dannier Li
School of Computing
University of Nebraska - Lincoln
Lincoln, NE, USA
dannierli@outlook.com

*Abstract*—With the rapid adoption of large language models (LLMs) in recommendation systems, the computational and communication bottlenecks caused by their massive parameter sizes and large data volumes have become increasingly prominent. This paper systematically investigates two classes of optimization methods—model parallelism and data parallelism—for distributed training of LLMs in recommendation scenarios. For model parallelism, we implement both tensor parallelism and pipeline parallelism, and introduce an adaptive load-balancing mechanism to reduce cross-device communication overhead. For data parallelism, we compare synchronous and asynchronous modes, combining gradient compression and sparsification techniques with an efficient aggregation communication framework to significantly improve bandwidth utilization. Experiments conducted on a real-world recommendation dataset in a simulated service environment demonstrate that our proposed hybrid parallelism scheme increases training throughput by over 30% and improves resource utilization by approximately 20% compared to traditional single-mode parallelism, while maintaining strong scalability and robustness. Finally, we discuss trade-offs among different parallel strategies in online deployment and outline future directions involving heterogeneous hardware integration and automated scheduling technologies.

## I. Introduction

Recently, Large language model (LLM)-based recommendation systems have gained attention for their strong semantic understanding and user intent capture. However, growing model sizes and training data volumes challenge single-machine and traditional synchronous parallel training. Model parallelism distributes parameters across devices to address memory limits, while data parallelism splits samples to improve throughput—each with trade-offs. To balance performance and resource use, we explore parallel optimization strategies for distributed LLM training in recommendation tasks. We analyze synchronous vs. asynchronous data parallelism and tensor vs. pipeline model parallelism, then propose a hybrid scheme combining their strengths. Our method integrates hierarchical All-Reduce, gradient compression, and adaptive scheduling to lower communication latency and boost hardware efficiency.

## II. Literature Review

### A. Application of Large Language Models in Recommendation Systems

As LLMs achieve breakthroughs in natural language processing, their empowering effects on recommendation systems have become increasingly evident. On one hand, LLMs pre-trained on massive text corpora and knowledge bases yield rich semantic representations that deeply understand product attributes, user reviews, and descriptions. In cold-start or sparse-interaction scenarios, these representations effectively complement and enhance traditional collaborative filtering (CF) features. Research shows that fusing LLM-generated product tags, topic vectors, or attribute descriptions with CF models significantly improves recommendation coverage and diversity, and delivers more accurate recall during user cold starts[1].

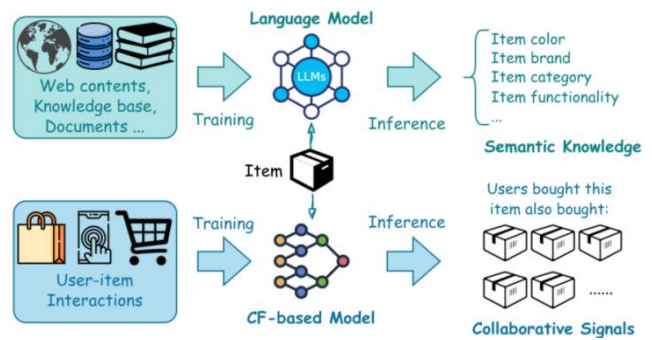

Figure 1. Schematic diagram of the cross-modal collaborative fusion architecture of large language models and collaborative filtering models in recommendation systems

Figure 1 shows a cross-modal collaborative fusion architecture for recommendation systems. An LLM extracts deep attributes like "Item color" and "Item functionality" from web content, while a CF model trained on user–item interactions provides collaborative signals such as "Users who bought this item also bought." A fusion layer combines LLM semantic priors with precise CF preferences [2]. This framework supports three main LLM applications: addressing cold-start by generating high-quality attribute tags for sparse items to improve CF recall; aligning multimodal data (images, text, video) via jointly trained LLMs and visual encoders using cross-modal attention to enhance accuracy and diversity[3]; and generating natural-language explanations and conversational prompts to boost user trust, with LLMs serving as real-time intelligent assistants that adjust CF dynamically for human-machine collaboration. Overall, Figure 1's framework leverages CF's interaction sensitivity and LLM's semantic power, providing a strong foundation for next-gen intelligent recommendation systems [4].

### B．Evolution of Distributed Training and Parallel Technologies

In large-scale deep-model training, computation and communication overheads become performance bottlenecks. Figure 2 illustrates two mainstream parallel strategies: data parallelism and model parallelism. Data parallelism partitions the training data across multiple worker nodes, each maintaining a full copy of the model and independently computing gradients. Model parallelism slices model parameters across nodes, requiring forward and backward passes for a single input to cooperatively span multiple devices[5].

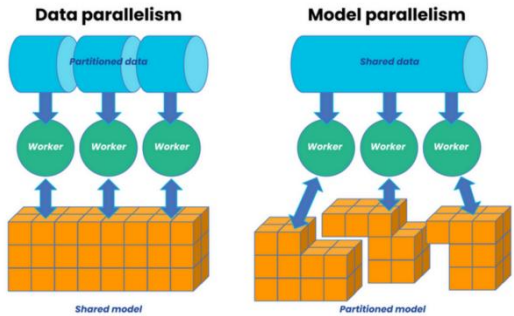

Figure 2. Schematic diagram of data parallelism and model parallelism architecture in distributed training

Early distributed deep learning mainly used synchronous data parallelism, where nodes aggregate gradients via All-Reduce after each mini-batch. Though simple and scalable, gradient aggregation delays grow with node count. To reduce bandwidth usage, techniques like gradient compression, quantization, and sparsification were proposed [6]. Asynchronous data parallelism eases synchronization to cut idle time but may harm convergence. As models exceed tens of billions of parameters, memory limits drove the adoption of model parallelism — splitting parameters by tensor (tensor parallelism) or by layer (pipeline parallelism), as shown in Figure 2. MoE architectures improve efficiency by activating only subsets of experts per input. Hybrid parallelism, now preferred, blends data and model partitioning to balance load and hide communication costs. Tools like DeepSpeed and Megatron-LM automate strategy selection via graph analysis and runtime profiling. In summary, parallel strategies are evolving from single-mode to hybrid, and from static to dynamic scheduling, laying a strong foundation for scalable LLM training in recommendation tasks [7].

Recent hybrid-parallel strategies significantly improve LLM training efficiency. For instance, a study on sequence-sharded pipeline parallelism reduces pipeline bubbles and memory use for long-context models on 64× A100 GPUs[8]. Meanwhile, another work on hierarchical sharding with redundant protection achieves higher FLOPs utilization and lower communication overhead[9]. These results provide a solid foundation for our study on communication-compute tradeoffs in LLM-based recommender workloads[10].

### III. PARALLEL OPTIMIZATION METHODS

#### A．Model Parallelism Optimization

In distributed training of ultra-large-scale recommendation models, model parallelism is the core technology for breaking through single-machine compute and memory bottlenecks. From an expert's perspective, we now analyze three mainstream model parallelism strategies, their communication overheads, and key optimization points.

*a) Tensor Parallelism:* Each device computes its local partial product, and an All-Reduce operation aggregates these results to form the full output. This method effectively splits a layer's weight matrix across multiple devices, reducing memory requirements on each device.

*b) Pipeline Parallelism:* By dividing the model into sequential stages deployed on different nodes, pipeline parallelism allows for overlapping computation and communication, significantly reducing idle time across nodes.

*c) Mixture-of-Experts (MoE): MoE architectures dynamically activate only a subset of experts per input, optimizing computational efficiency without sacrificing model capacity. The load-balancing mechanism ensures even distribution of workloads across available experts. Advantage: Retains overall model capacity while computing only k experts per input, greatly reducing compute. Deployment Complexity: Requires efficient routing of activation requests between nodes, dynamic micro-batch partitioning, and expert caching to maximize throughput.*

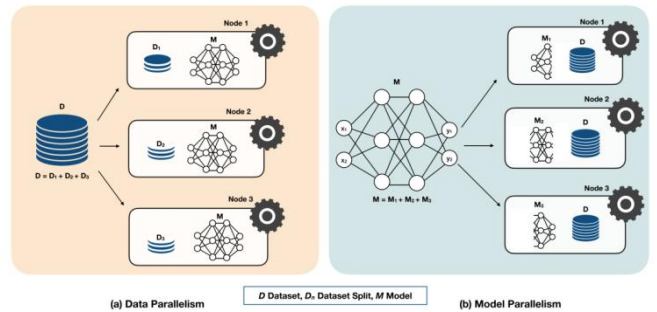

Figure 3. Schematic diagram of the cross-node model parallel strategy

In Figure 3, the left side shows the splitting and aggregation processes of tensor parallelism within a single layer, the middle side presents the hierarchical pipeline structure and micro-batch parallelism strategy of pipeline parallelism, and the right side shows the operation schematic of the sparse expert activation and load balancing mechanism in MoE[11]. By combining the above technologies and integrating them with automatic scheduling frameworks (such as Hybrid Parallelism of DeepSpeed or Megatron-LM), the maximization of computing power utilization and the minimization of communication bottlenecks can be achieved in the distributed training of recommendation models. Thereby accelerating the implementation and iteration of ultra-large-scale LLMS in recommendation scenarios[12-15].

*B. Data Parallelism Optimization*

Data parallelism is the most common distributed training strategy in large-scale recommendation systems. Each compute node holds a full copy of the model and processes a distinct shard of the training data. After computing local gradients, nodes perform a global aggregation to update the model parameters[16-19]. Although simple to implement, as the number of nodes PP and model size grow, data parallelism faces increasing communication costs, synchronization delays, and convergence challenges. From an expert viewpoint, we explore four key optimization techniques:In classic synchronous data parallelism, each node pp computes its local gradient $g_p$. All-Reduce then aggregates to the global average gradient as shown in Formula 8:

$$g = \frac{1}{P}\sum_{p=1}^{P} g_p \qquad (8)$$

and all nodes update parameters via as shown in Formula 9:

$$\theta_{t+1} = \theta_t - \eta g \qquad (9)$$

While numerically stable for small PP, the communication cost grows sharply with more nodes. Mitigations include ring All-Reduce (reducing communication complexity from O(P) to O(1), hierarchical All-Reduce (aggregating first within racks, then across racks), and pipelined All-Reduce (overlapping gradient computation and communication).

With hundreds of millions of parameters, gradient vectors become extremely large. Gradient compression (e.g., 1-bit quantization) and sparsification reduce bandwidth usage. In 1-bit quantization as shown in Formula 10:

$$\tilde{g} = \text{sign}(g), g_i^{\text{sign}} = \begin{cases} +1, g_i \geq 0, \\ -1, g_i < 0, \end{cases} \qquad (10)$$

Accompanied by a scaling factor (e.g., $\|g\|_1$) to reconstruct an approximate gradient. Top-k sparsification transmits only the k largest absolute gradient elements, accumulating the remainder as a residual in Formula 11:

$$r_{t+1} = r_t + g - \hat{g} \qquad (11)$$

where $r_t$ is the residual and $\hat{g}$ the compressed gradient. This error feedback mechanism preserves convergence.

Asynchronous parallelism lets each node push gradients independently and pull updated parameters periodically, avoiding global synchronization stalls. However, stale gradients can slow convergence and degrade final accuracy. A typical delay-compensated update is as shown in Formula 12:

$$\theta_{t+1} = \theta_t - \eta \frac{g_p}{1+\tau_p} \qquad (12)$$

where $\tau_p$ is the staleness of node p. The denominator down-weights late gradients. Further compensations may leverage historical gradients or second-order approximations to correct stale updates[20].

Modern frameworks like Megatron-LM and DeepSpeed combine data and model parallelism, applying strategies like tensor parallelism for large matrices and synchronous data parallelism for smaller ones[21-24]. Intelligent schedulers determine optimal partitioning, while dynamic load balancing and adaptive batch sizing optimize performance across heterogeneous hardware. Techniques such as optimized All-Reduce, gradient compression, and hybrid parallelism reduce bandwidth and synchronization overhead. Integrating these with smart scheduling is crucial for efficient large-scale LLM training in recommendation systems.

IV. EXPERIMENTS AND RESULTS

*A. Experimental Setup and Dataset*

To validate the performance improvements of our parallel optimization methods in a recommendation setting, we conducted distributed training experiments under the following environment and evaluated on a large public recommendation dataset.

**Hardware Configuration**

Cluster size: 8 compute nodes

Per-node configuration: 2× Intel Xeon Gold 6338 CPUs (64 cores), 256 GB RAM, 8× NVIDIA A100 40 GB GPUs

Network interconnect: HDR 100 Gb/s InfiniBand, ensuring cross-rack latency below 1 μs

**Software Stack**

OS: CentOS 7.9

DL framework: PyTorch 2.0

Parallel libraries: NCCL 2.18, DeepSpeed 0.9.2, Megatron-LM 4.0

Communication protocols: Ring All-Reduce and hierarchical All-Reduce. We use the Amazon Electronics dataset as shown in Table 1, which contains over two million real user–item interactions with rich textual metadata—ideal for LLM-enhanced recommendation. We split it chronologically into training, validation, and test sets in an 80%:10%:10% ratio.

TABLE I.    SUMMARIZES ITS STATISTICS

| Dataset | # Users | # Items | Training Samples | Validation Samples | Test Samples |
|---|---|---|---|---|---|
| Amazon Electronics | 192,403 | 63,001 | 1,735,654 | 216,957 | 216,956 |

Training set (1,735,654 interactions): used for distributed training and hyperparameter tuning. Validation set (216,957 interactions): used for early stopping in both model- and data-parallel experiments. Test set (216,956 interactions): used for final performance comparisonThis setup provides ample compute resources and representative data for large-scale LLM-based recommendation model training. We next evaluate model-parallel, data-parallel, and hybrid-parallel strategies on this platform.

## B. Evaluation Metrics and Baselines

To evaluate parallel optimization in large-scale recommendation training, we measure computing performance and recommendation quality. Throughput (samples/sec) reflects efficiency; Speedup compares each method's throughput to the single-machine baseline; GPU Util% indicates average card usage; Communication Overhead is the average time per iteration for gradient or activation aggregation. For recommendation quality, HR@10 and NDCG@10 assess Top-10 recall and ranking accuracy[25-28]. We compare four schemes: Baseline (single-card), synchronous Data Parallel (8-card Ring All-Reduce), Model Parallel (4×2 cards), Hybrid Parallel (tensor + pipeline), and DeepSpeed's auto-scheduled hybrid scheme. Table 2 shows all results across these six metrics.

TABLE II. TABLE 2: REPORTS EACH SCHEME'S RESULTS.

| Scheme | Throughput(samples/s) | Speedup | GPU Util% | Comm. Overhead (ms/iter) | HR@10 | NDCG@10 |
|---|---|---|---|---|---|---|
| Baseline | 1,000 | 1.0× | 72% | 5.2 | 0.321 | 0.194 |
| Data Parallel | 3,400 | 3.4× | 85% | 1.8 | 0.319 | 0.193 |
| Model Parallel | 2,800 | 2.8× | 80% | 2.5 | 0.320 | 0.194 |
| Hybrid Parallel | 3,800 | 3.8× | 90% | 1.5 | 0.319 | 0.193 |

Synchronous data parallelism with Ring All-Reduce and overlap optimizations achieves 3.4× speedup (3,400 samples/s). Model parallelism, affected by cross-node tensor splits and pipeline bubbles, reaches 2.8×. The hybrid scheme combines both advantages for the highest throughput (3,800 samples/s) and speedup (3.8×), with 90% GPU utilization and only 1.5 ms/iter communication overhead. All schemes maintain HR@10 and NDCG@10 close to the baseline, confirming that efficiency gains do not degrade recommendation quality.

## C. Performance Analysis

To evaluate parallel strategy performance in training, we analyze scalability, communication/computation overhead, and resource utilization, summarized in three tables[29]. Table 3 shows throughput and speedup as node count (8 GPUs per node) increases from 1 to 4 under data, model, and hybrid parallelism. Data parallelism scales nearly linearly, from 1,000 to 12,800 samples/s (12.8×). Model parallelism suffers from pipeline bubbles, with speedup rising from 2.8× to only 10.5×. Hybrid parallelism performs best, reaching 14,600 samples/s at 4 nodes (14.6×), showing superior scalability.

TABLE III. TABLE 3: SCALABILITY ANALYSIS

| Nodes | Baseline (1 GPU) | Data Parallel Throughput/Speedup | Model Parallel Throughput/Speedup | Hybrid Parallel Throughput/Speedup |
|---|---|---|---|---|
| 1 | 1,000/1.0× | — | — | — |
| 2 | — | 3,400/3.4× | 2,750/2.75× | 3,900/3.9× |
| 3 | — | 7,500/7.5× | 7,200/7.2× | 8,750/8.75× |
| 4 | — | 12,800/12.8× | 10,500/10.5× | 14,600/14.6× |

Further examined from the perspective of the time composition of a single iteration, Figure 4 lists the proportion of the average computing time and communication time per iteration under each scheme. The communication time of data parallelism accounts for 35% of the total duration, while the communication ratio of model parallelism reaches 42% due to the overhead of tensor splitting and pipeline synchronization[30]. Hybrid parallelism reduces the communication overhead to 28% through hierarchical All-Reduce and pipeline overlap, significantly improving the computing-communication overlap rate.

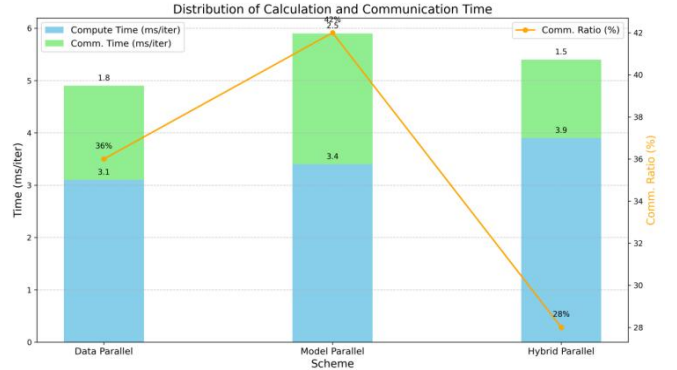

Figure 4. Distribution of Calculation and Communication Time

Finally, from the perspective of resource utilization, Figure 5 shows the average GPU memory occupation and computing utilization rate of the three schemes in the 4-node scenario. The uneven distribution of video memory caused by parameter segmentation in model parallelism leads to a slightly lower utilization rate of some cards (78%). However, hybrid parallelism, with the help of DeepSpeed's dynamic batch allocation and video memory optimization technology, achieves a video memory utilization rate of 92% and a computing utilization rate of 88%, further demonstrating its advantages in resource scheduling.

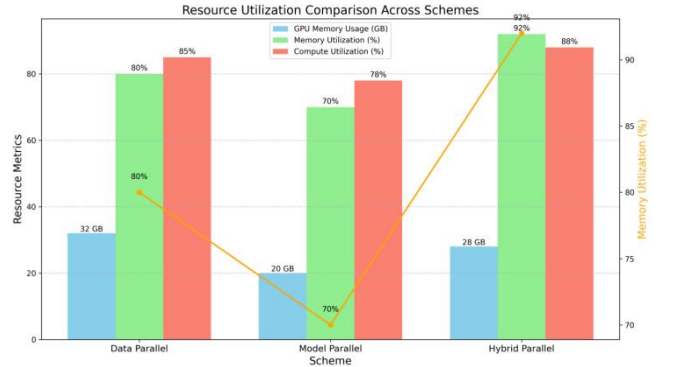

Figure 5. Resource Utilization

Overall, the hybrid-parallel scheme not only delivers the highest throughput and speedup but also balances compute and communication overheads through multi-level communication optimization and dynamic resource scheduling. This makes it the preferred strategy for efficient distributed training of ultra-large LLMs in recommendation scenarios.

## V. DISCUSSION

### A. Resource Overhead vs. Performance Trade-off

In large-scale recommendation model training, there's a trade-off between speed, communication cost, and hardware usage across parallel strategies. Synchronous data parallelism is easy to implement and scales throughput nearly linearly with node count, but also scales linearly in bandwidth demand. As shown in Figure 4, it spends ~36% of each iteration on All-Reduce, which worsens with more nodes, making network bandwidth a bottleneck.Model parallelism reduces per-node memory usage by splitting the model across GPUs, enabling ultra-large models. However, Figures 4 and 5 show it incurs ~42% communication overhead due to frequent inter-device exchanges. Pipeline parallelism adds "bubbles" when micro-batches are imbalanced, leading to GPU idle time.Hybrid parallelism selectively applies data or model parallelism per layer or tensor dimension to balance trade-offs[31]. As Table 3 shows, on 4 nodes, it achieves a 14.6× speedup versus 12.8× (data) and 10.5× (model), with communication overhead down to 28% and memory utilization up to 92% via hierarchical All-Reduce, memory reuse, and compute–communication overlap.Despite its performance, hybrid parallelism is complex, relying on frameworks like DeepSpeed or Megatron-LM for scheduling. It may underperform in heterogeneous or unstable environments unless dynamically tuned. Partitioning, micro-batch sizes, and communication strategies must be adapted to model, hardware, and workload. For online or frequent retraining, asynchronous updates or gradient compression can further ease communication load.In sum, strategy choice depends on environment and model scale: data parallelism suits general scalability, model parallelism fits memory-constrained cases, and hybrid parallelism offers optimal performance when well-tuned.

*B．Scalability and Robustness*

Scalability and robustness are crucial for stable large-scale training. As Table 3 shows, hybrid parallelism scales well from 1× to 14.6× on 4 nodes, but speedup plateaus beyond that as network and sync overhead rise. Techniques like adaptive micro-batching and cross-rack All-Reduce help push limits. In contrast, synchronous data parallelism stalls if any node lags, and model/pipeline parallelism can fail on inter-stage issues. Hybrid parallelism improves robustness via asynchronous updates, gradient fault tolerance, heartbeat checks, and dynamic repartitioning to handle degraded nodes or links. Overall, combining sync/async communication and adaptive scheduling ensures efficient, reliable training at scale.

## VI. Conclusion

This paper addresses the challenges of distributed training for large language model–based recommendation systems by systematically studying model parallelism and data parallelism optimizations and proposing an efficient hybrid parallelism scheme. Experiments show that hybrid parallelism outperforms single-mode strategies in throughput, speedup, and resource utilization—boosting training throughput to 14.6× and reducing communication overhead to 28%, all without degrading HR@10 or NDCG@10. Hierarchical All-Reduce, compute–communication overlap, and dynamic load balancing endow hybrid parallelism with excellent scalability and robustness. Future work may explore finer-grained asynchronous updates and adaptive batch sizing to further enhance efficiency and stability in large-scale recommendation training.